\documentclass{svproc}
\usepackage[utf8]{inputenc}
\usepackage[english]{babel}

\usepackage{url}

\usepackage{textcomp}

\usepackage[sectionbib]{natbib}
\bibpunct{(}{)}{;}{a}{}{,}%
\usepackage{graphicx}
\usepackage{amsmath,amssymb}
\usepackage{multirow}
\usepackage{longtable}

\begin{document}
\mainmatter 
\title{The Largest Russian Optical Telescope BTA: Current Status and Modernization Prospects} 

\titlerunning{The Largest Russian Optical Telescope BTA} 

%
\author{D.~O.~Kudryavtsev \and V.~V.~Vlasyuk} 

\authorrunning{Kudryavtsev \& Vlasyuk} 

\tocauthor{Dmitry O. Kudryavtsev, Valery V. Vlasyuk } 

\institute{Special Astrophysical Observatory, Russian Academy of Sciences, Nizhny Arkhyz, Russia, \email{dkudr@sao.ru}} 

\maketitle 

\begin{abstract} The Russian 6-m telescope (BTA), once the largest telescope in the world and now the largest optical telescope in Russia, has been successfully operating for almost 45 years. In this paper, we briefly overview the observing methods the instrument can currently provide, the ongoing projects on the development of scientific equipment, the status of the telescope among the world`s and Russian astronomical communities, our ambitions to attract new users, and the prospects the observatory wishes to realize in the near future. 
\keywords{telescopes} 
\end{abstract} 

\section{Introduction} 

At the end of this year, the largest national optical telescope, Big Telescope Azimuthal (BTA), with a 6-meter primary mirror marks its 45th anniversary, counting from the date of signature of the certificate by the interagency acceptance committee on December 30, 1975. Figure~1 shows the copies of the first images that were obtained with the BTA and confirmed the compliance of its parameters with the design characteristics. Until 1993, the BTA had been the largest telescope in the world, now it remains among the top twenty largest instruments. It is still the country`s largest telescope and a joint-use instrument for Russian and foreign scientists. 

	\begin{figure}
				\includegraphics[width=\columnwidth]{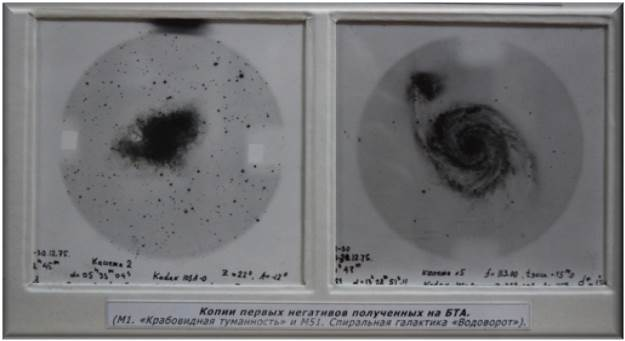}
		\caption{The copies of the first images obtained with the BTA.}
	\end{figure}

During its operation, the telescope has been constantly developed and upgraded. In the first decades, the main manufacturers of the telescope, among them LOMO (Leningrad Optical Mechanical Association), LZOS (Lytkarino Optical Glass Plant), and GOI (Vavilov State Optical Institute), had been providing assistance and patronage; recently, this task has almost completely fallen on the SAO~RAS personnel. Among the main results that contributed to maintaining a high level of research at the BTA, we can highlight the creation of the new automated control system (BTA ACS), introduction of remote observations for most scientific equipment units, implementation of a program for the creation of modern light detectors, and, finally, the development of observing methods that cover the entire spectrum of scientific problems solved with the telescope. 

Summing up the intermediate results, we should also note a number of our failures, among those a technological error in the primary mirror repolishing procedure and the lack of methods for observing in the infrared range at the BTA. Among the reasons that had a negative impact are insufficient funding for complex technological projects, technological gap in the Russian industry, and the deficit of human resources. 

Rather a modest place of our telescope in the current world rankings, new large ground-based projects expected in this decade---the Extremely Large Telescope (39.3-m mirror, consortium of European countries) and the Giant Magellan Telescope (25.4~m, consortium led by the USA)---force us to search for a niche where our telescopes are going to be in demand and competitive at the world level. 

It can be argued that in recent decades the BTA has not been able to compete with new-generation world telescopes in addressing most challenging observational problems. In addition to purely technological reasons---the absence of active and adaptive optics on the telescope in the presence of a ``thick'' glass mirror with an average, by contemporary standards, diameter---there is the astroclimate, which is dramatically inferior to the best foreign sites both in seeing and in the number of clear nights. 

Nevertheless, the search for our ``ecological niche'' does not seem to us a hopeless venture. The BTA remains a fairly large instrument, it has a unique geographical location in latitude and longitude among the world`s telescopes. Not least, the telescope is equipped and is constantly being reequipped with modern scientific instruments and light detectors, which allows astronomers to obtain world-class scientific results. In addition, an essential function of SAO~RAS is the preservation of personnel potential, the staff of highly qualified specialists having experience in creating original research and observing methods and their implementation at a large telescope. 

\section{Scientific Equipment and Methods of Observation} 

The history of the observatory and the BTA is inextricably associated with development of observing methods and scientific equipment. We can say with confidence that the current observing methods used at the telescope are organically related to the scientific interests of SAO~RAS scientists, and thus already reflect BTA characteristics and naturally denote a niche for our research. It can be designated as (1)~conducting research by classical means: high and low resolution spectroscopy and photometry of objects difficult to access for other large instruments and (2)~the use of technically sophisticated and not widely used methods: high-accurate spectropolarimetry, fast photometry, panoramic spectroscopy, and speckle interferometry. Of course, the development and support of such methods should be continued in the future.

\begin{figure}
				\centerline{\includegraphics[width=0.45\columnwidth]{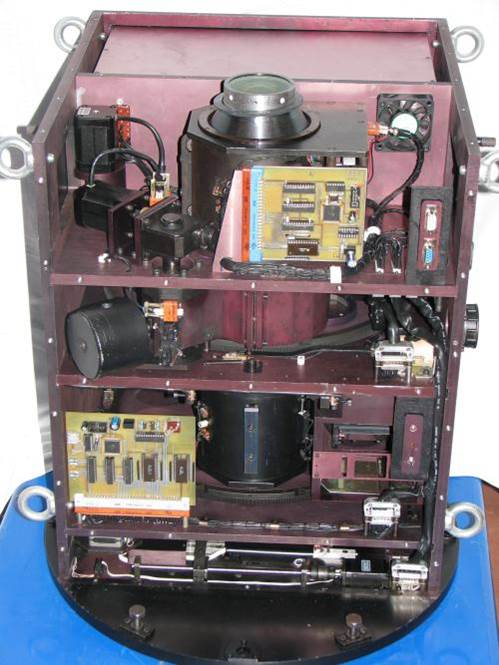}}
		\caption{The universal spectrograph SCORPIO-2.}
	\end{figure}

\subsection{Spectroscopy and Photometry of Faint Objects} 

The multipurpose spectrographs of the BTA primary focus: SCORPIO-1 \citep{sco1} and SCORPIO-2, developed in the Laboratory of Spectroscopy and Photometry of Extragalactic Objects under the leadership of Chief Researcher V.\,L.~Afanasiev,
implement classical methods of spectroscopy and photometry of faint objects in a $6'\times6'$ field of view in the whole visible wavelength range (from 360 to 1000~nm) with resolution $R=\lambda/\Delta\lambda$ from 500 to 2500. SCORPIO-2 (shown in Fig.~2) perform in addition highly accurate polarimetric and spectropolarimetric observations \citep{sco2_pol} and is equipped with a set of scanning Fabry--Perot interferometers \citep{sco2_fpi2002, sco2_fpi2008, sco2_fpi2015} with resolution $R$ in the interval from 250 to 16\,000, allowing panoramic spectroscopy in the same wide field ($6'\times6'$) but in a narrow spectral range (smaller than 10~nm). The developers also intend to introduce an integral-field spectroscopy unit \citep{sco2_ifu}, which is going to implement spectroscopy of extended objects with angular resolution better than $1''$ and  spectral resolution $R$ in the range of 1000--2800 within a $16''\times16''$ field. SCORPIO-1 and SCORPIO-2 are equipped with large-format light detectors based on CCDs up to $2048\times4600$~pixels, manufactured in recent years by the Advanced Design Laboratory (ADLab) of SAO RAS.

\subsection{Stellar Spectroscopy} 

The list of scientific equipment designed to obtain stellar spectra at the BTA with resolution greater than 10\,000 is represented by two instruments: the Nasmyth Echelle Spectrograph (NES), created under the leadership of V.\,E.~Panchuk in the Astrospectroscopy Laboratory, and the Main Stellar Spectrograph (MSS), the only instrument that remained in the BTA equipment out of those created by LOMO and GOI during the construction of the telescope. 

	\begin{figure}
				\includegraphics[width=\columnwidth]{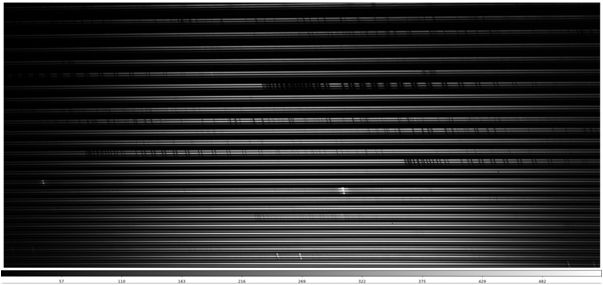}
		\caption{A planetary nebula spectrum obtained with NES in the 5700--8500\,\AA\ range.}
	\end{figure}

NES \citep{nes} obtains spectra of stars up to $10^m$ with\linebreak \mbox{$R\sim40\,000$--$50\,000$} and a simultaneously recorded spectral range of about\linebreak 300~nm. To increase efficiency, the spectrograph has 3-fragment image slicers with slit widths of $0\farcs4$ and $0\farcs6$, there are also classical ($0\farcs6\times2\farcs5$) and adjustable \mbox{($0\farcs2$--$2''\times0''$--$40''$)} slits. The instrument is equipped with two interchangeable cross-dispersers for operation in the $\lambda$ ranges of 305--600~nm and 480--1000~nm. Figure~3 shows an example of a NES spectrum obtained in the near infrared wavelength range.

The Main Stellar Spectrograph has been repeatedly upgraded by V.\,E.~Pan\-chuk, G.\,A.~Chountonov, and I.\,D.~Naidenov \citep{mss} in order to install large-sized diffraction gratings, introduce CCD systems, and implement spectropolarimetry (circular polarization analyzers). The spectrograph obtains spectra of stars brighter than $12^m$ with resolution up to $R\sim15\,000$ and simultaneously recorded ranges of 55~nm for wavelengths shorter than 500 nm and more than 80~nm for the 560-900~nm range. The instrument operates mainly in the spectropolarimetric mode with a circular polarization analyzer combined with an image slicer producing 7~fragments for each polarization. 

Both spectrographs are equipped with large-format light detectors developed in SAO~RAS ADLab based on the $2048\times4608$~px E2V CCD42-90 chips. 

\subsection{Speckle Interferometry with the Diffraction-Limited Angular~Resolution} 

The Group of High Resolution Methods in Astronomy led by Yu.\,Yu.~Balega has been developing and constantly improving the speckle interferometry method \citep{speckle} since the 80s. The method is aimed at studying binary and multiple star systems, single stars, asteroids, and other celestial bodies with an angular resolution close to the diffraction limit of the telescope. This can be achieved by Fourier analysis of interference patterns from sub-apertures of the primary mirror, detected with exposures of 5--30 ms (the time of ``freezing'' of atmospheric fluctuations). The instrument operates in the 500-900~nm range with a set of narrow interference filters, its field of view with the currently used $512\times512$~px Andor EMCCD iXon Ultra 897 light detector is 4\farcs4, 7\farcs1, or 28\farcs2 depending on a microlense used, the limiting magnitude is $15^m$, the angular resolution is $0\farcs02$. 

\subsection{Superfast Photometry} 

For studies of optical variability of astrophysical objects with high temporal resolution (up to 1~ms), i.e., stellar mass black hole candidates, pulsars, flare stars, etc., the Group of Relativistic Astrophysics  under the leadership of Leading Researcher G.\,M.~Beskin designed the Multimode Panoramic Photopolarimeter \citep{mppp}. The method can use two types of detectors depending on required temporal resolution: either two position-sensitive detectors (to achieve a temporal resolution of the order of 1~ms) or an EMCCD camera with charge amplification (for times of the order of 0.1~s). The field of view is $1'$ in the $UBVR$ panoramic photometry mode, there is also a slit mode with adjustable slit widths and heights \mbox{($0''$--$10''\times10''$--$60''$)}. Linear polarization measurements (using a double Wollaston prism) and spectral modes with an Abbe prism and a diffraction grating are available.

\subsection{Light Detectors} 

The Advanced Design Laboratory began development of optical light detectors for telescopes in the 80s under the leadership of Ph.D. in Engineering Science S.\,V.~Markelov, and since the beginning of the 90s, as technical characteristics improved, the detectors began to be introduced into regular BTA observations, gradually replacing, where possible, photographic plates and detectors based on image intensifiers. Improvement of the quality of semiconductor crystals and development of the signal processing technique allowed us by the turn of the century to move from $256\times256$~px systems with quantum efficiency of 30\%--40\%\ in the red part of the spectrum and additional noises of the order of hundreds of electrons to $2048\times2048$~px systems with quantum efficiency of almost 90\%\ and noises of several electrons. 

For the last 15~years, SAO telescopes have almost everywhere used systems sized from $2048\times2048$~px to greater formats, and all further development followed mainly the path of increasing the stability of signal detection, expanding the spectral range to the red part, and eliminating effects of light interference in detector layers. To date, the observatory has mastered the production of detection systems based on the E2V $4096\times4096$ chips, which have almost ideal properties: high quantum efficiency and a low level of interference noise in the near IR range. As part of joint work with Russian enterprises,  ADLab has developed a technology for creating mosaic detection systems of an almost arbitrary format limited only by price and the size of the cryostat. Figure~4 shows a CCD system designed to equip the BTA fiber-fed spectrograph. 

\begin{figure}
				\centerline{\includegraphics[width=0.45\columnwidth]{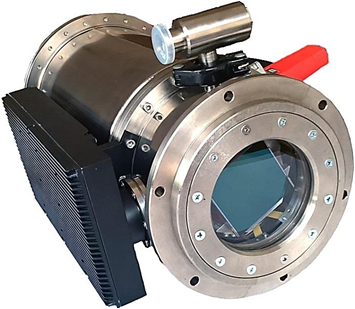}}
		\caption{A CCD system based on a $4128\times4112$~px E2V CCD 231-84 detector, created for the BTA fiber-fed spectrograph, was accepted for observations in the summer of 2020.}
	\end{figure}

\section{BTA Projects under Development} 

\renewcommand{\labelitemi}{\textbullet} 

Among the equipment at various stages of implementation, the appearance of which is expected in the near future, we can name: \begin{itemize} 

\item Fiber-fed optic spectrograph for the BTA \citep{ffs}. It was manufactured as part of the SAO RAS project ``Evolution of Stars from Their Birth to the Origin of Life'' supported by the Russian Science Foundation in 2014--2018. The spectrograph is installed on the telescope and equipped with a light detector, test observations are carried out with a temporary optical camera, first results have been obtained, the manufacturer is adjusting the permanent optical camera. 

\item Photospectropolarimeter in the Nasmyth-1 focus. The instrument is intended for operating in the semi-automatic mode for observations of alert transients. A $1024\times1024$~px Andor EMCCD system with subsecond temporal resolution is planned as a detector. 

\item To increase the recording rate in panoramic observations at the BTA,\linebreak SAO~RAS ADLab started in 2019 the development of systems based on CMOS sensors within the project ``Development of the Large Unique Scientific Facility Big Telescope Alt-Azimuthal'' supported by the Ministry of Science and Higher Education in 2019-2020 within the federal target program ``Research and Development in Priority Areas of Development of the Scientific and Technological Complex of Russia for 2014--2020.\!'' The laboratory have created a prototype of a fast photodetector based on the CMOS sensor Gpixel GSENSE4040CMN and developed design documentation for a photodetector based on the astronomical-class large-format back-illuminated sensor GSENSE6060. 

\item The BTA Multiobject Fiber Spectrograph (MOFS) is being upgraded. It is planned to develop a new camera, improve the technology for making masks, upgrade the control computer and software of the instrument. 

\item The observatory is planning to introduce the first, experimental, adaptive system in the BTA Nasmyth-2 focus, which should improve the seeing in observations with high spectral resolution methods. The purchased adaptive system for correction of large-scale wavefront aberrations will perform aberration correction up to the 4th order at a frequency of 200~Hz with an accuracy of $\lambda/15$. 

\item The work on upgrading the BTA Echelle Spectropolarimeter for Primary Focus (ESPriF) continues; the instrument is planned to be equipped with a $2048\times2048$~px Andor iKon-L 936 light detector. 
\end{itemize}	

\section{The BTA As a Unique Scientific Facility}	

The status of Russia's ``Unique Scientific Facilities'' held by the BTA and\linebreak \mbox{RATAN-600} of SAO RAS has in recent years become increasingly important in maintaining the provision of regular observations and financial stability of the observatory as a whole. Before the reform of the state academies of sciences in Russia, this status often remained nominal without significantly affecting the terms of funding. With a new regulatory framework and after the transfer of economic and administrative management functions from the Russian Academy of Sciences to the jurisdiction of the Federal Agency for Scientific Organizations in 2013, the situation with shared research facility centers (SRFCs) and the unique scientific facilities (USFs) has drastically changed. 

For instance, the basic funding of SAO RAS is calculated without taking into account the needs of the telescopes, as for an average scientific institute, and even allowing for the regional location (which, for example, reduces the estimated amount of the salary fund). In this situation, the USF status and the presence of external users make it possible to request and receive additional funds for operation of the telescopes, as well allowing participation in project competitions of the Ministry of Science and Higher Education to attract additional funding for development and renovation of scientific equipment. The loss of the USF status will inevitably lead to the loss of funds for operation of the telescopes. 

Among the main target indicators established by a special decree of the Russian Government on the requirements for SRFCs and USFs are the equipment workload (the ratio of the actual USF operating time to the maximum possible operating time per year), which must be 70\% at least, the presence of external users, the number of organizations applying for a USF, and publications in indexed journals. Therefore, SAO RAS is extremely interested in maintaining high efficiency of the telescopes and attracting external users. In 2019, the BTA workload was 96\%, the workload in the interests of external users 36\%, we had 13~applying organizations, 41 papers were published based on observations obtained with the telescope in the journals indexed in the Web of Science. 

Even though it should be said that the principles of financing the science in Russia are far from ideal---for example, the specific character of development and production of unique scientific equipment is poorly taken into account, most project competitions are aimed at instant scientific return, amid significant investments in ensuring the work of scientists, inadequate attention is paid to the support of scientific and technical personnel---in general, the government aims at development of the scientific sphere, invests heavily in science and even, to some extent, provides feedback communication, which gives hope for gradual improvement of policy in this area. 

\section{The BTA in the SAO RAS Development Program until~2024} 

In conclusion, we present some points of the SAO RAS development program for 2019--2023 created in the framework of the implementation of the national ``Science'' project. Surely, the percentage of completion of these plans will be determined by the availability of funding, but the program shows well the directions along which SAO RAS scientists are going to progress in the coming years. The observatory supposes to develop the existing and create new effective methods for dealing most topic astronomical problems: studying the large-scale structure of the Universe, active galactic nuclei, detecting fast radio bursts, identifying optical transients associated with gamma-ray bursts and gravitational events, and exoplanet research. In particular, the development of the BTA complex involves the following projects: 
\begin{itemize} 

\item development of new scientific instruments to increase technical capabilities of the BTA and expand the observed electromagnetic range: a near-infrared spectrograph for the 0.8--2.5 {\textmu}m wavelength interval and new-generation 2D spectrographs and spectropolarimeters for the primary focus; 

\item development of new high-speed large-format light detectors and mosaic systems for optical telescopes; 

\item development of adaptive optics for partial compensation of atmospheric distortions and vibrations of the telescope mount; 

\item creation of a system for photometric control of the reflective layer in applying the coating onto the BTA primary mirror. 
\end{itemize} 

\section*{Acknowledgements} Observations with the SAO RAS telescopes are supported by the Ministry of Science and Higher Education of the Russian Federation (including agreement No.\,05.619.21.0016, project~ID RFMEFI61919X0016). The renovation of telescope equipment is currently provided within the national ``Science'' project. The authors would like to thank the Russian Foundation for Basic Research for the consistent support of the RAS large telescope user conferences and the Russian Telescope Time Allocation Committee sessions during many years up to 2018 (e.g., RFBR projects 17-02-20559, 18-02-20136). We should also appreciate the support from the Russian Science Foundation, which allows the observatory to develop new scientific equipment, particularly the BTA fiber-fed spectrograph (RSF project No.~14-50-00043).


\begin{thebibliography}{11}
\expandafter\ifx\csname natexlab\endcsname\relax\def\natexlab#1{#1}\fi

\bibitem[{{Afa\-na\-si\-ev} \& {Moiseev}(2005)}]{sco1}
{Afa\-na\-si\-ev}, V.~L. \& {Moiseev}, A.~V. 2005, Astronomy Letters, 31, 194

\bibitem[{{Afanasiev} \& {Amirkhanyan}(2012)}]{sco2_pol}
{Afanasiev}, V.~L. \& {Amirkhanyan}, V.~R. 2012, Astrophysical Bulletin, 67,
  438

\bibitem[{{Afanasiev} {et~al.}(2018){Afanasiev}, {Egorov}, \&
  {Perepelitsyn}}]{sco2_ifu}
{Afanasiev}, V.~L., {Egorov}, O.~V., \& {Perepelitsyn}, A.~E. 2018,
  Astrophysical Bulletin, 73, 373

\bibitem[{{Maksimov} {et~al.}(2009){Maksimov}, {Balega}, {Dyachenko},
  {Malogolovets}, {Rastegaev}, \& {Semernikov}}]{speckle}
{Maksimov}, A.~F., {Balega}, Y.~Y., {Dyachenko}, V.~V., {et~al.} 2009,
  Astrophysical Bulletin, 64, 296

\bibitem[{{Moiseev}(2002)}]{sco2_fpi2002}
{Moiseev}, A.~V. 2002, Bulletin of the Special Astrophysical Observatory, 54,
  74

\bibitem[{{Moiseev}(2015)}]{sco2_fpi2015}
{Moiseev}, A.~V. 2015, Astrophysical Bulletin, 70, 494

\bibitem[{{Moiseev} \& {Egorov}(2008)}]{sco2_fpi2008}
{Moiseev}, A.~V. \& {Egorov}, O.~V. 2008, Astrophysical Bulletin, 63, 181

\bibitem[{{Panchuk} {et~al.}(2014){Panchuk}, {Chuntonov}, \& {Naidenov}}]{mss}
{Panchuk}, V.~E., {Chuntonov}, G.~A., \& {Naidenov}, I.~D. 2014, Astrophysical
  Bulletin, 69, 339

\bibitem[{{Panchuk} {et~al.}(2017){Panchuk}, {Klochkova}, \& {Yushkin}}]{nes}
{Panchuk}, V.~E., {Klochkova}, V.~G., \& {Yushkin}, M.~V. 2017, Astronomy
  Reports, 61, 820

\bibitem[{{Plo\-khot\-ni\-chen\-ko} {et~al.}(2009){Plo\-khot\-ni\-chen\-ko},
  {Beskin}, {de Bur}, {Karpov}, {Bad'in}, {Lyubetskaya}, {Lyubetskij}, \&
  {Pavlova}}]{mppp}
{Plo\-khot\-ni\-chen\-ko}, V.~L., {Beskin}, G.~M., {de Bur}, V.~G., {et~al.}
  2009, Astrophysical Bulletin, 64, 308

\bibitem[{{Valyavin} {et~al.}(2020){Valyavin}, {Musaev}, {Perkov}, {Aitov},
  {Bychkov}, {Drabek}, {Shergin}, {Sazonenko}, {Kukushkin}, {Galazutdinov},
  {Emelyanov}, {Yakopov}, {Burlakova}, {Bertaux}, {Tavrov}, {Korablev},
  {Yushkin}, {Valeev}, {Gadelshin}, {Kim}, {Han}, \& {Lee}}]{ffs}
{Valyavin}, G.~G., {Musaev}, F.~A., {Perkov}, A.~V., {et~al.} 2020,
  Astrophysical Bulletin, 75, 191

\end{thebibliography}
\end{document}